\documentclass[iop]{emulateapj}

\def\gtorder{\mathrel{\raise.3ex\hbox{$>$}\mkern-14mu
             \lower0.6ex\hbox{$\sim$}}} 
\def\ltorder{\mathrel{\raise.3ex\hbox{$<$}\mkern-14mu
             \lower0.6ex\hbox{$\sim$}}}
\def\ltsima{$\; \buildrel < \over \sim \;$}
\def\simlt{\lower.5ex\hbox{\ltsima}}
\def\gtsima{$\; \buildrel > \over \sim \;$}
\def\simgt{\lower.5ex\hbox{\gtsima}} 

\shorttitle{SN2003ie}
\shortauthors{Arcavi et al.}

\begin{document} 


\title{Supernova 2003ie Was Likely a Faint Type IIP Event}


\author{Iair~Arcavi\altaffilmark{1}$^{,}$\altaffilmark{2},
Avishay~Gal-Yam\altaffilmark{1}
and Sergey G. Sergeev\altaffilmark{3}
}

\altaffiltext{1}{Department of Particle Physics and Astrophysics, The Weizmann Institute of Science, Rehovot 76100, Israel}
\altaffiltext{2}{iair.arcavi@weizmann.ac.il}
\altaffiltext{3}{Crimean Astrophysical Observatory, P/O Nauchny, Crimea 98409, Ukraine}



\newpage

\begin{abstract} 

We present new photometric observations of Supernova (SN) 2003ie starting one month before discovery, obtained serendipitously while observing its host galaxy. With only a weak upper limit derived on the mass of its progenitor ($<25M_{\odot}$) from pre-explosion studies, this event could be a potential exception to the ``red supergiant (RSG) problem'' (the lack of high mass RSGs exploding as Type~IIP supernovae). However, this is true only if SN2003ie was a Type~IIP event, something which has never been determined. Using recently derived core collapse SN light curve templates, as well as by comparison to other known SNe, we find that SN2003ie was indeed a likely Type~IIP event. However, it is found to be a member of the \emph{faint} Type~IIP class. Previous members of this class have been shown to arise from relatively low mass progenitors ($<12M_{\odot}$). It therefore seems unlikely that this SN had a massive RSG progenitor. The use of core collapse SN light curve templates is shown to be helpful in classifying SNe with sparse coverage. These templates are likely to become more robust as large homogeneous samples of core collapse events are collected.

\end{abstract} 




\section{Introduction} 

Core collapse supernovae (SNe) mark the end stage of massive (M$>8M_{\odot}$) stars. They are largely divided into two groups: H-poor explosions (Type~I SNe) and H-rich explosions (Type~II SNe; see Filippenko 1997 for a review). Several additional subclasses have been identified, including interacting SNe~IIn (see e.g. Schlegel 1990, Kiewe et al. 2012 and references therein), linearly declining SNe IIL, plateau light curve SNe IIP, H-deficient SNe IIb and events displaying long rise times similar to SN1987A. Arcavi et al. (2012) find a distinct subdivision in the $R$-band light curves of the IIP, IIL and IIb subtypes.

While the origin of Type~IIL events is currently under debate (e.g., Kasen \& Bildsten 2010, Woosley 2010, Moriya \& Tominaga 2012), SNe IIP have been robustly associated with the explosions of red supergiants (RSGs) using direct pre-explosion imaging of the SN sites (see Smartt 2009, Leonard 2011 for reviews; Van Dyk et al. 2012, Fraser et al. 2012 for recent results). However, while RSGs in the local group have masses up to $\sim30M_{\odot}$ (Levesque et al. 2005), Smartt et al. (2009) find that none of the observed progenitors of confirmed SNe~IIP are more massive than $\sim 17M_{\odot}$. It is unclear why no RSG with a mass in the range $17-30M_{\odot}$ has been observed to explode as a Type~IIP event,  and this discrepancy has been dubbed the ``RSG problem''. 

Assuming a Salpeter Initial Mass Function (IMF), Smartt et al. (2009) find that the probability of their sample not including any SN from a massive RSG by chance is very low ($1.8\%$). The IMF alone is therefore not enough to resolve this problem. Massive RSGs might explode as SNe IIL, IIb or IIn, but so far these SNe were found to arise from lower mass progenitors (Smartt et al. 2003), binaries (Podsiadlowski et al. 1993; Woosley et al. 1994; Maund et al. 2004) and luminous blue variables (Gal-Yam \& Leonard 2009), respectively. Other options could be that massive RSGs are enshrouded by thick dust, making their explosions difficult to detect, or that they collapse directly to black holes.

One possible exception to the RSG problem is SN2003ie, for which only a loose progenitor mass upper limit ($25M_{\odot}$) could be derived. However, due to the sparse data available for this SN, its subclass could not be determined. Determining whether SN2003ie was a Type~IIP event or not could have implications for the RSG problem, thus motivating this study.

SN2003ie was discovered on 2003 September 19 (Arbour \& Boles 2003; UT times used throughout). A spectrum taken by Benetti et al. (2003) revealed it was a Type~II SN. Harutyunyan et al. (2008) find that the spectrum was similar to that of SN1998A. Both SN2003ie and SN1998A show a significant ($\sim2100$km/s) blueshift in their Halpha emission line. This prompted the suggestion that SN2003ie, like SN1998A, is a member of the slowly rising SN1987A-class. These events are thought to arise from blue supergiants (BSGs; Kleiser et al. 2011, Pastorello et al. 2012, Taddia et al. 2012). Indeed SN1987A had a massive BSG progenitor ($16-22M_{\odot}$; Arnett et al. 1989). If this were the case, SN2003ie would not be of any relevance to the RSG problem. However, Pastorello et al. (2005) note that blueshifted Halpha emission has also been observed in the Type~IIP SN1999em. It is therefore not clear which subclass of Type II events SN2003ie belongs to.

Here we present $BVRI$ photometry of SN2003ie obtained serendipitously during observations of its host galaxy, NGC 4051. Our data cover approximately 130 days starting one month prior to discovery. We compare the light curve of this event to those of other SNe in order to better constrain its subtype.

\section{Observations}
\begin{figure}
\includegraphics[width=85mm]{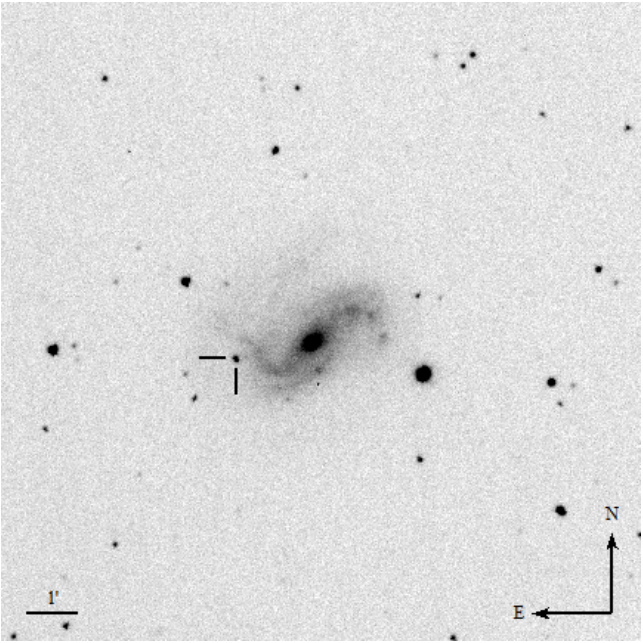}
\caption{An $R$-band image of SN2003ie in the galaxy NGC 4051, taken with the 70-cm Crimean Reflector on 2003 August 18.}
\label{disco_image}
\end{figure}

\begin{figure}
\includegraphics[width=85mm]{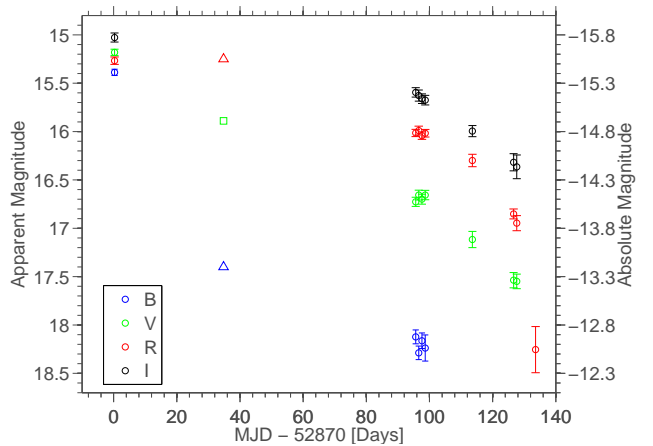}
\caption{The $BVRI$ light curves of SN2003ie. Observations taken on the same day are averaged. The circles denote our photometry, the square denotes synthetic photometry derived from the Benetti et al. (2003) spectrum and the triangles denote lower limits for the synthetic photometry derived using this spectrum (which only partially covered the $B$ and $R$ bands).}
\label{lc}
\end{figure}

We used the the AP7p CCD camera mounted on the 70-cm telescope at the Crimean Astrophysical Observatory to observe the galaxy NGC 4051 during the months of July 2003 to January 2004 (see Segeev et al. 2005 for more information on the instrument). Several observations captured SN2003ie, starting from 2003 August 18 (approximately one month before discovery; Figure \ref{disco_image}). We employ the mkdifflc routine (Gal-Yam et al. 2004, 2008) within IRAF\footnote{IRAF (Image Reduction and Analysis Facility) is distributed by the National Optical Astronomy Observatories, which are operated by AURA, Inc., under cooperative agreement with the National Science Foundation.} to subtract the host galaxy using pre-explosion images taken with the same setup. We calibrate our photometry to Sloan Digital Sky Survey (SDSS) stars in the field, converting their magnitudes to the Johnson-Cousins system using the Jordi et al. (2006) equations (see Kiewe et al. 2012 for more details). We correct for Galactic extinction using the Schlafly \& Finkbeiner (2011) recalibration of the Schlegel, Finkbeiner \& Davis (1998) infrared-based dust map, retrieved via the NASA Extragalctic Database (NED\footnote{NED is operated by the Jet Propulsion Laboratory, California Institute of Technology, under contract with the National Aeronautics and Space Administration.}). We adopt a distance modulus of 30.8 mag to the SN (taken from NED). Our photometry is presented in Table \ref{phot} and the light curve is shown in Figure \ref{lc}. We also include synthetic photometry derived from the classification spectrum taken on 2003 September 22 by Benetti et al. (2003) and presented in Harutyunyan et al. (2008). Because the spectrum did not fully cover the wavelength range of the $B$ and $R$ filters, we derive only lower limits on the synthetic luminosity in these bands. 

We constrain the explosion time to be between 2003 July 19 and 2003 August 18. Our photometry and the Benetti et al. (2003) spectrum are digitally released via WISeREP\footnote{http://www.weizmann.ac.il/astrophysics/wiserep} (Yaron \& Gal-Yam 2012).

\begin{table}
\scriptsize
\caption{Photometry of SN2003ie.}
\begin{tabular}{lllll}
\hline
\hline
MJD & B & V & R & I \tabularnewline
\hline    
52870.282 & $15.83$ ($0.07$) & $15.19$ ($0.06$) & $15.25$ ($0.08$) & $15.02$ ($0.09$)\tabularnewline
52870.285 & $15.27$ ($0.07$) & $15.18$ ($0.07$) & $15.35$ ($0.08$) & $15.11$ ($0.10$)\tabularnewline
52870.288 & $15.32$ ($0.06$) & $15.24$ ($0.07$) & $15.28$ ($0.08$) & $15.05$ ($0.09$)\tabularnewline
52870.291 & $15.33$ ($0.06$) & $15.26$ ($0.07$) & $15.31$ ($0.08$) & $15.00$ ($0.09$)\tabularnewline
52904.781\footnote{Dervied from the Benetti et al. (2003) spectrum.} & $>17.40$ & $15.89$ & $>15.25$ & \tabularnewline
52965.642 & $18.15$ ($0.15$) & $16.71$ ($0.11$) & $16.05$ ($0.08$) & $15.56$ ($0.10$)\tabularnewline
52965.644 & $18.10$ ($0.12$) & $16.82$ ($0.09$) & $16.01$ ($0.07$) & $15.62$ ($0.10$)\tabularnewline
52965.646 & $18.09$ ($0.18$) & $16.77$ ($0.09$) & $16.05$ ($0.07$) & $15.61$ ($0.09$)\tabularnewline
52965.648 & $18.35$ ($0.13$) & $16.75$ ($0.09$) & $16.07$ ($0.08$) & $15.66$ ($0.10$)\tabularnewline
52966.602 & $18.32$ ($0.13$) & $16.59$ ($0.11$) & $15.98$ ($0.11$) & $15.66$ ($0.10$)\tabularnewline
52966.604 & $18.26$ ($0.14$) & $16.66$ ($0.10$) & $16.02$ ($0.09$) & $15.62$ ($0.12$)\tabularnewline
52966.605 & $18.33$ ($0.13$) & $16.83$ ($0.10$) & $16.03$ ($0.08$) & $15.71$ ($0.12$)\tabularnewline
52966.607 & $18.44$ ($0.15$) & $16.69$ ($0.09$) & $16.05$ ($0.08$) & $15.59$ ($0.12$)\tabularnewline
52967.617 & $18.01$ ($0.17$) & $16.89$ ($0.09$) & $16.09$ ($0.08$) & $15.67$ ($0.10$)\tabularnewline
52967.619 & $18.13$ ($0.11$) & $16.71$ ($0.11$) & $16.10$ ($0.08$) & $15.73$ ($0.10$)\tabularnewline
52967.621 & $18.29$ ($0.17$) & $16.59$ ($0.10$) & $16.06$ ($0.08$) & $15.59$ ($0.09$)\tabularnewline
52967.623 & $18.42$ ($0.19$) & $16.75$ ($0.11$) & $16.04$ ($0.08$) & $15.72$ ($0.10$)\tabularnewline
52968.647 & $18.08$ ($0.19$) & $16.72$ ($0.10$) & $16.03$ ($0.08$) & $15.63$ ($0.09$)\tabularnewline
52968.649 & $18.61$ ($0.42$) & $16.65$ ($0.09$) & $16.08$ ($0.08$) & $15.67$ ($0.10$)\tabularnewline
52968.651 & $18.17$ ($0.18$) & $16.66$ ($0.09$) & $16.04$ ($0.08$) & $15.70$ ($0.11$)\tabularnewline
52968.653 & $18.29$ ($0.21$) & $16.74$ ($0.11$) & $16.04$ ($0.08$) & $15.77$ ($0.10$)\tabularnewline
52983.589 &  & $17.11$ ($0.15$) & $16.33$ ($0.12$) & $16.13$ ($0.11$)\tabularnewline
52983.591 &  & $17.16$ ($0.15$) & $16.31$ ($0.13$) & $15.95$ ($0.12$)\tabularnewline
52983.592 &  & $17.24$ ($0.14$) & $16.32$ ($0.13$) & $15.98$ ($0.11$)\tabularnewline
52983.594 &  & $17.10$ ($0.22$) & $16.36$ ($0.13$) & $15.99$ ($0.13$)\tabularnewline
52996.630 &  & $17.42$ ($0.14$) & $16.94$ ($0.09$) & $16.41$ ($0.19$)\tabularnewline
52996.632 &  & $17.62$ ($0.16$) & $16.87$ ($0.13$) & $16.18$ ($0.15$)\tabularnewline
52996.634 &  & $17.66$ ($0.20$) & $16.88$ ($0.09$) & $16.43$ ($0.21$)\tabularnewline
52996.636 &  & $17.59$ ($0.13$) & $16.84$ ($0.09$) & $16.32$ ($0.15$)\tabularnewline
52997.629 &  & $17.66$ ($0.16$) & $17.04$ ($0.17$) & $16.56$ ($0.31$)\tabularnewline
52997.631 &  & $17.45$ ($0.14$) & $17.01$ ($0.15$) & $16.30$ ($0.25$)\tabularnewline
52997.634 &  & $17.68$ ($0.14$) & $16.78$ ($0.14$) & $16.42$ ($0.25$)\tabularnewline
52997.636 &  & $17.55$ ($0.15$) & $17.08$ ($0.17$) & $16.25$ ($0.14$)\tabularnewline

\hline
\label{phot}
\end{tabular}
\normalsize
\end{table}

\section{Discussion}

\begin{figure*}
\includegraphics[width=18cm]{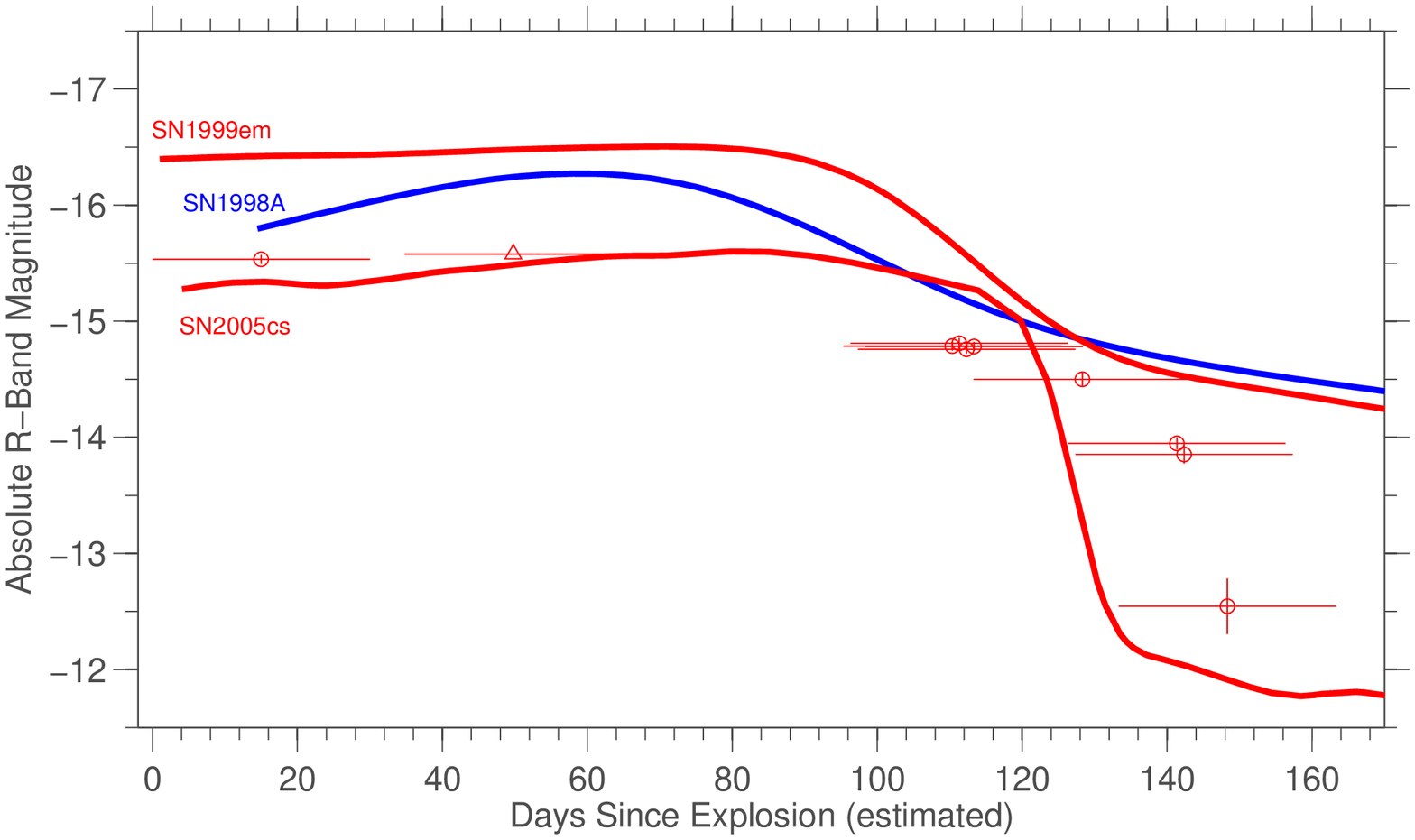}
\includegraphics[width=18cm]{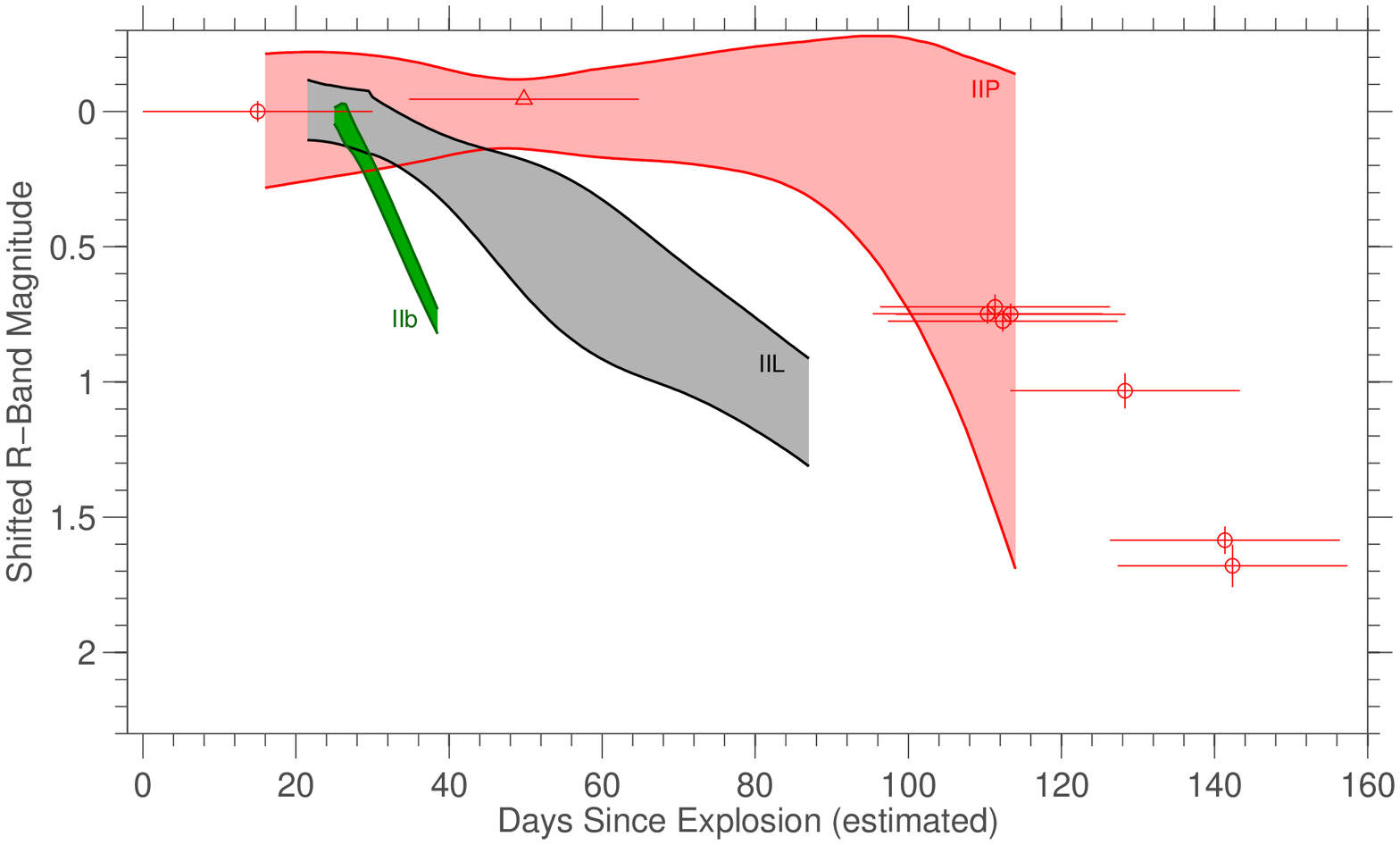}
\caption{Top Panel: $R$-band light curves of SN2003ie (same as in Fig \ref{lc}), the 87A-like SN1998A (Pastorello et al. 2005) and the SNe~IIP 1999em (Leonard et al. 2002) \& 2005cs (Pastorello et al. 2009). Spline fits were applied to the reference SNe. Despite being similar to SN1998A spectroscopically (Harutyunyan et al. 2008), SN2003ie is not photometrically an 87A-like event. Rather it is more similar to faint SNe IIP such as SN2005cs, albeit with a more gradual termination of the plateau. The uncertainty in explosion time is marked by the horizontal $30$-day error bars (note that this is a systematic uncertainty).
Bottom Panel: The $R$-band light curve of SN2003ie (shifted from Fig \ref{lc}) in the context of the normalized SN~II templates presented in Arcavi et al. (2012). Despite the gradual plateau end, the light curve of SN2003ie is most consistent with the Type~IIP template. The uncertainty in explosion time is marked by the horizontal $30$-day error bars (note that this is a systematic uncertainty).}
\label{comparisons}
\end{figure*}

We compare the $R$-band light curves of SN2003ie, SN1998A and the Type~IIP SNe 1999em and 2005cs in figure \ref{comparisons}. It is clear that photometrically, SN2003ie is not a member of the 87A-class, especially when considering the drop in brightness approximately $100$ days after our first detection. This behaviour, together with the observed $B$-band decline (Fig. \ref{lc}), is typical of Type~IIP events. The possiblity of a massive BSG progenitor for SN2003ie is therefore disfavoured. 

While the drop from the plateau for SN2003ie is more gradual than some other SNe~IIP, it is still far from the Type~IIL and Type~IIb light curve templates identified by Arcavi et al. (2012; bottom panel of our Fig. \ref{comparisons}). 

SN2003ie is thus a Type~IIP SN, likely arising from an RSG progenitor which can not be constrained to a mass $<17M_{\odot}$.

However, the absolute plateau magnitude of SN2003ie puts it in the faint SN~IIP category. Several other faint Type~IIP SNe had relatively low mass progenitors (see Fraser et al. 2011 and references therein). Therefore, we conclude that this event is not likely to have been the explosion of a high mass RSG.

\section{Summary}

We have presented new photometric observations of SN2003ie, which strongly suggest it was a faint Type~IIP event. Despite the spectroscopic similarities to SN1998A (Harutyunyan et al. 2008) we show that these two events are not similar photometrically. We use the templates derived by Arcavi et al. (2012) to also rule out a IIL or IIb classification. 

With only an upper limit of $25M_{\odot}$ on its progenitor mass, SN2003ie could thus have been the exception to the red supergiant problem. However, as a member of the faint Type~IIP subclass, this option is not likely.

Our work demonstrates that using core collapse SN light curve templates could prove usefull for transient identification when only sparse photometric and spectroscopic data is available. Constructing robust templates, however, requires a large homogeneous sample of well classified events. Such a sample will likely be produced using current and future wide field transient surveys, such as the Palomar Transient Factory (Law et al. 2009, Rau et al. 2009).

A.G. and I.A. acknowledge support by the Israeli and German-Israeli Science Foundations, and an EU/FP7/ERC grant.

\end{document}